\documentclass{webofc}
\usepackage[varg]{txfonts}   
\usepackage{graphicx}
\usepackage{dcolumn}
\usepackage{bm}
\usepackage{ifpdf}
\usepackage{graphicx}
\usepackage{amssymb}
\usepackage{epsfig}
\usepackage{enumerate}
\usepackage{pifont}
\usepackage{verbatim}
\usepackage{lineno}
\usepackage{float}
\usepackage{appendix}
\usepackage{listings}

\newcommand{\shlong}{\ensuremath{\sigma^{2}_{\rm long}}}

\newcommand{\Jpsi} {\mbox{J\kern-0.05em /\kern-0.05em$\psi$}\xspace}

\newcommand{\sN}{$\sqrt{s_{\rm NN}}$}
\newcommand{\pT}{\ensuremath{p_{\rm T}}\xspace}

\newcommand{\zT}{\ensuremath{z_{\rm T}}}

\newcommand{\ptIsoCh}{\ensuremath{p_{\rm T}^{\rm iso,~ch}}}

\newcommand{\isogamma}{$\gamma^{\rm iso}$\xspace}

\newcommand {\mom}    {\mbox{\rm  GeV$\kern-0.15em /\kern-0.12em c$}}
\newcommand {\gmom}   {\mbox{\rm  GeV$\kern-0.15em /\kern-0.12em c$}}
\newcommand {\mass} {\mbox{\rm  GeV$\kern-0.15em /\kern-0.12em c^2$}}

\begin{document}
\title{Direct photon \pT~spectra and correlations measured with ALICE}
%
%

\author{\firstname{Carolina} \lastname{Arata}\inst{1}\fnsep\thanks{\email{carolina.arata@cern.ch}} on behalf of the ALICE Collaboration 
}

\institute{Laboratoire de Physique Subatomique et de Cosmologie, Université Grenoble-Alpes, CNRS-IN2P3, Grenoble, France}

\abstract{
        This contribution discusses the measurements of direct photons in pp and Pb--Pb collisions from the LHC Run 2, as recorded by the ALICE experiment. Specifically, we focus on the isolated photons results obtained at \sN~= 5.02 TeV. The isolated photons \isogamma~spectra in pp and Pb--Pb collisions are presented together with the nuclear modification factor, that is found to be consistent with unity. The azimuthal correlations of \isogamma~with hadrons in Pb--Pb and the hadron \zT~distributions $D(\zT)$ are presented, showing a centrality-dependent suppression compared to the pQCD NLO pp reference.}
\maketitle
\section{Introduction}
\label{intro}
Measurements of direct photons provide valuable information on the properties of the quark-gluon plasma (QGP) formed in ultra-relativistic collisions of heavy nuclei, because they are colour-neutral.
Direct photons are considered as those photons not originated from hadronic decays.
They can be distinguished in: photons from Compton ($\rm qg\rightarrow\gamma q$) and annihilation ($\rm q\bar{q}\rightarrow\gamma g$) processes produced at the initial hard scattering, the so-called $\gamma_{2\rightarrow2}$; photons from parton fragmentation and photons produced during all the phases of the heavy-ion collision.
The direct photons production, obtained using the decay-photon subtraction method \cite{WA98:2000ulw}, measured in Pb--Pb collisions by ALICE has been illustrated in the presentation and more details can be found in this proceeding \cite{Marin:2023kqi}. Therefore, the following proceeding will focus on the $\gamma_{2\rightarrow2}$ measured via the isolation method \cite{ALICE:2019rtd, Ichou:2010wc}.
Photons from 2$\rightarrow$2 processes, dominating the direct photon yield at \pT~$\gtrsim$ 5 GeV/$c$, are characterized by the absence of event activity in their vicinity from the hard process. Hence, selecting such "isolated photons" can suppress part of the photons from hadronic decays and parton fragmentation $\gamma_{\text{fragm}}$.
Isolated photon measurements in pp and Pb--Pb collisions can constrain NLO pQCD predictions and PDFs and nPDFs, respectively.
Since photons do not interact strongly, hadrons correlated with \isogamma~are a promising channel to study the energy loss in heavy-ion collisions and to constrain the $Q^2$ of the initial hard scattering, obtaining information on the amount of energy lost by the parton recoiling off the photon.

\section{Analysis details}\label{sec-1}
The pp and Pb--Pb collisions data for these measurements were collected at the LHC during Run 2 with the ALICE detector \cite{ALICE:2008ngc, ALICE:2014sbx} at the center of mass energy per nucleon pair \sN~= 5.02 TeV. The charged particles (tracks) are reconstructed by the Inner Tracking System (ITS) in pp collisions and by combining the ITS and the Time Projection Chamber (TPC) in Pb--Pb collisions. Photons are detected using the Electromagnetic Calorimeter (EMCal). From the particle energy deposit in the calorimeter cells, neighbouring cells can be grouped into a cluster, whose shape can be quantified by the elongation parameter \shlong~(see \cite{ALICE:2022qhn}). 
A selection based on \shlong~allows to distinguish between single clusters, from 2$\rightarrow$2 photons, $N_{\rm narrow}$ (0.10~<~\shlong~<~0.30) and elongated clusters from neutral meson decays, $N_{\rm wide}$, like $\pi^{0}\rightarrow\gamma\gamma$ (0.40~<~\shlong~<~2.00). Neutral meson two photon decays tend to merge into a single cluster for $E_{\pi^{0}}$ > 6 GeV due to the Lorentz boost.
The purity of the direct $\gamma_{2\rightarrow2}$ sample can be enhanced applying an isolation criterion.
The photon is declared isolated (denoted with superscript "iso") when in a cone with $R~=~\sqrt{(\varphi-\varphi^{\gamma})^2+ (\eta-\eta^{\gamma})^2}$, the sum of the transverse momentum of charged particles inside the cone (\ptIsoCh), corrected by the collision underlying event (UE) contribution in the cone $\rho_{\rm UE}$, is smaller than a value, as shown in equation \ref{eq-1}.
\begin{equation}
        \label{eq-1}
        \ptIsoCh~=~\sum p_{\rm T}^{\rm~tracks~in~cone} - \rho_{\rm UE}~\pi~R^{2}       
\end{equation}
The $\rho_{\rm UE}$ is not correlated to the hard process at the origin of the direct photon and it is estimated as the sum of the tracks~\pT~outside the isolation cone but within the same $\varphi$ covered by the cone. The \ptIsoCh~has to be below 1.5 GeV/$c$ and a radius $R$ of 0.2 is used. In the spectra analysis, a radius $R$ = 0.4 is also tested.
A quantity of isolated decay and fragmentation photons still contributes to the sample of \isogamma. The purity is estimated using the ABCD method \cite{ALICE:2019rtd}. In pp collisions it increases from around 0.1 at 12 GeV/$\it{c}$ up to a plateau around 0.4 for a \pT~above 20 GeV/$\it{c}$, while in central (peripheral) Pb--Pb collisions it increases from around 0.3 (0.2) up to around 0.6 (0.5). The purity increases from pp towards central PbPb collisions partly due to the increasing neutral mesons suppression. 
\section{Results}
\label{sec-2}
\begin{figure}[h]
     \centering
     \includegraphics[width=4.6cm,clip]{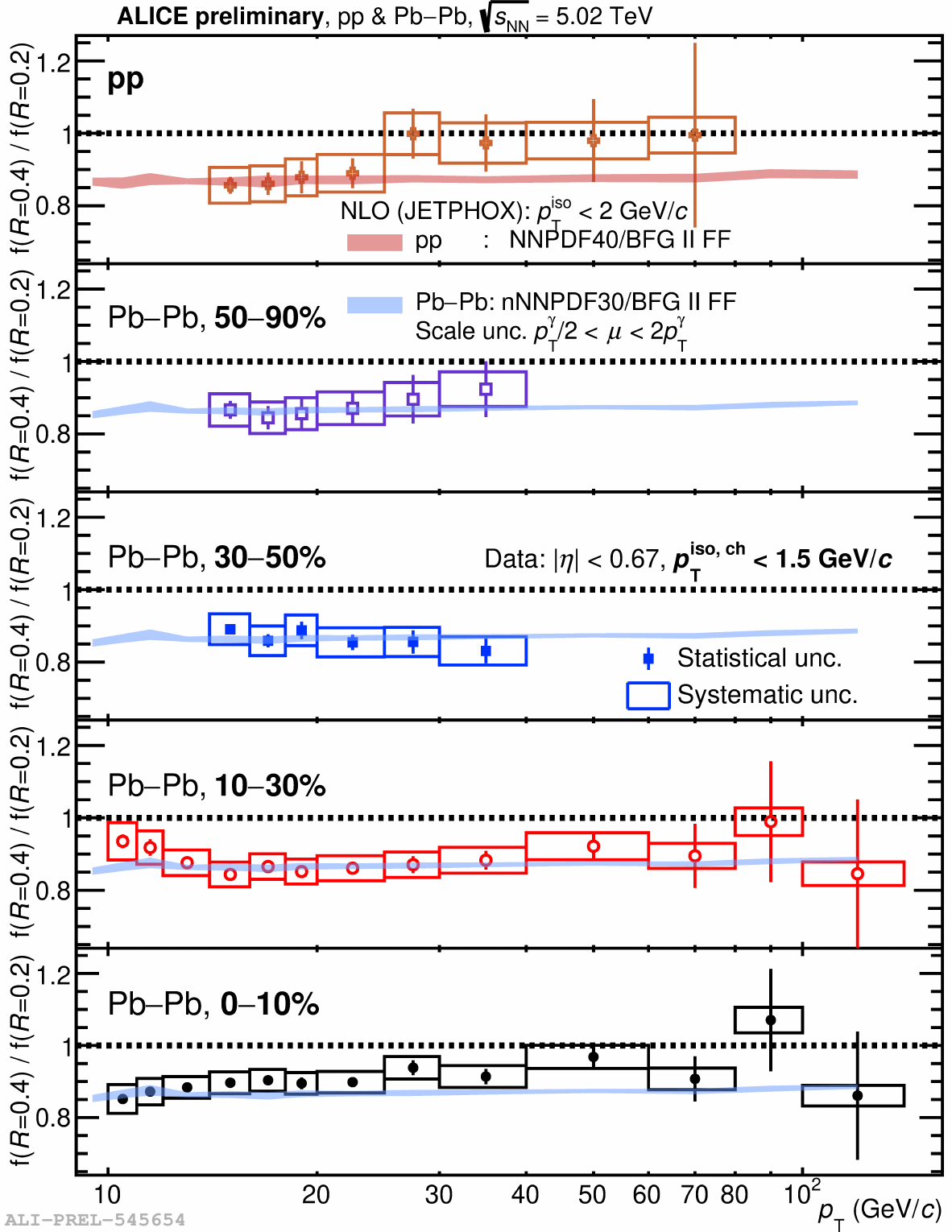}
     \includegraphics[width=6.6cm,clip]{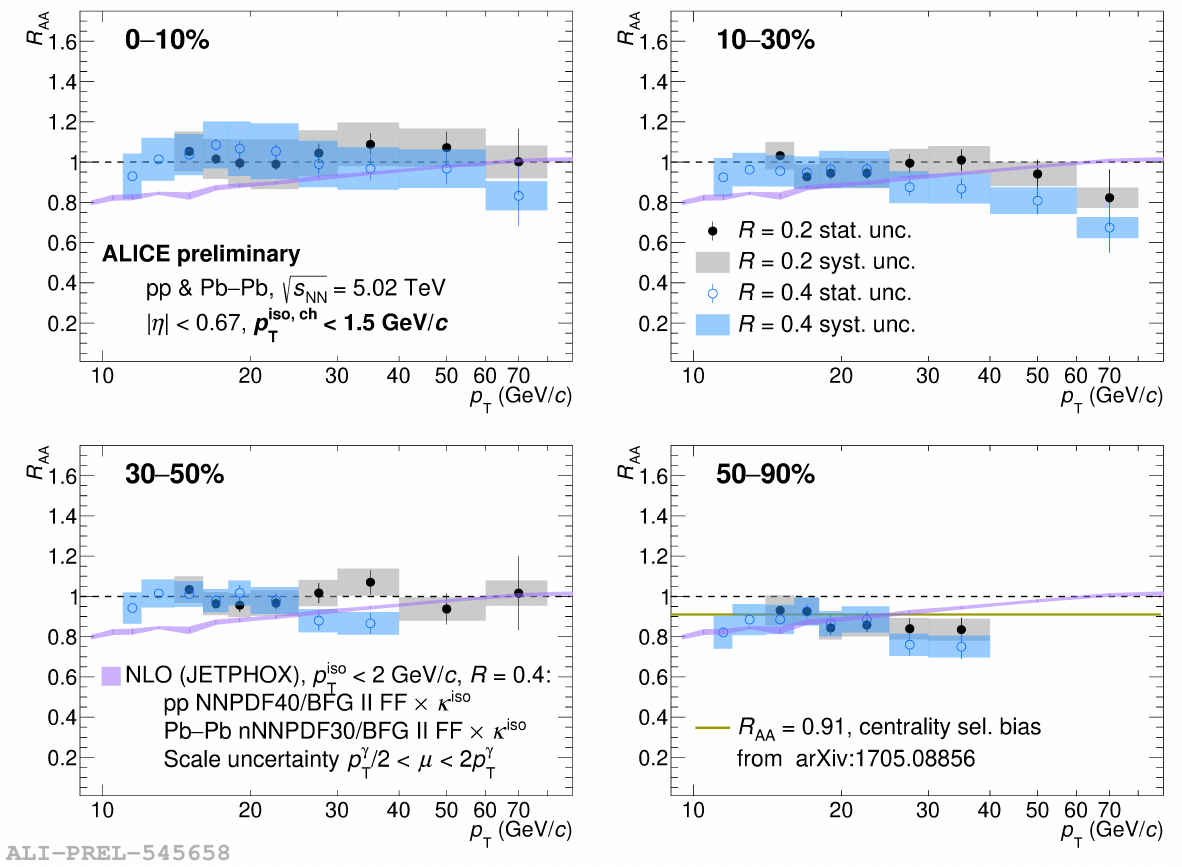}
     \caption{Left: ratio of cross sections with different $R$ as a function of \pT, from pp (top panel) to central Pb--Pb collisions (bottom panel) compared with JETPHOX NLO calculations in red (light blue) for pp (Pb--Pb) collisions. Right: $R_{\text{AA}}$ for different centralities obtained with a radius $R$ of 0.2 (0.4) in black (in light blue) compared with JETPHOX NLO calculations in violet.}
     \label{fig-1}       
\end{figure}

The \isogamma~cross sections in pp and in Pb--Pb collisions at \sN~=~5.02 TeV are measured and compared to NLO calculations using JETPHOX \cite{Catani:2002ny} with BFG II fragmentation functions and with NNPDF40 \cite{NNPDF:2021njg} for pp collisions and nNNPDF30 \cite{AbdulKhalek:2022fyi} for Pb--Pb collisions. The theory is centrality independent. Good agreement is found with theory for both radii and collisional systems.
From the two cross sections with different $R$, it is possible to compute the cross sections ratio $\text{f}(R\text{=0.4})/\text{f}(R\text{=0.2})$, shown in Figure~\ref{fig-1}-left. This ratio is sensitive to the fraction of $\gamma_{\text{fragm}}$ surviving the isolation. The dominant uncertainty at low~\pT~is the systematic uncertainty from the UE subtraction. The comparison with NLO calculations shows quite good agreement. The nuclear modification factor $R_{\text{AA}}$ as a function of \pT~is shown in Figure~\ref{fig-1}-right.
\par The $R_{\text{AA}}$ is compatible with unity, as expected. In peripheral collisions, the result is still in agreement with unity within the uncertainties but shifted down to a lower value due to a centrality selection bias of the Glauber model that causes an apparent suppression.
This deviation is known and the model described in Ref. \cite{Loizides:2017sqq} predicts a shift to 0.91, in agreement with the measurement. The dominant uncertainty at low~\pT~is the systematic uncertainty from the isolation probability. 
The data have been compared to the corresponding ratio of NLO calculations: there is an agreement with the NLO pQCD ratio above 20 GeV/$c$, whereas below there are some tensions due to the difference between PDF and nPDF.
This measurement is compatible with CMS results \cite{CMS:2020oen} for both cone radii, and extends the LHC measurements towards lower \pT. 
As the $R_{\text{AA}}$ indicates, the photon yield is not modified by the QGP, hence, the \isogamma~measurement can be used to tag the initial energy of the parton emitted in the opposite azimuthal direction in the 2$\rightarrow$2 process. 
\par
\begin{figure}[h]
  \centering
  \includegraphics[width=4.2cm,clip]{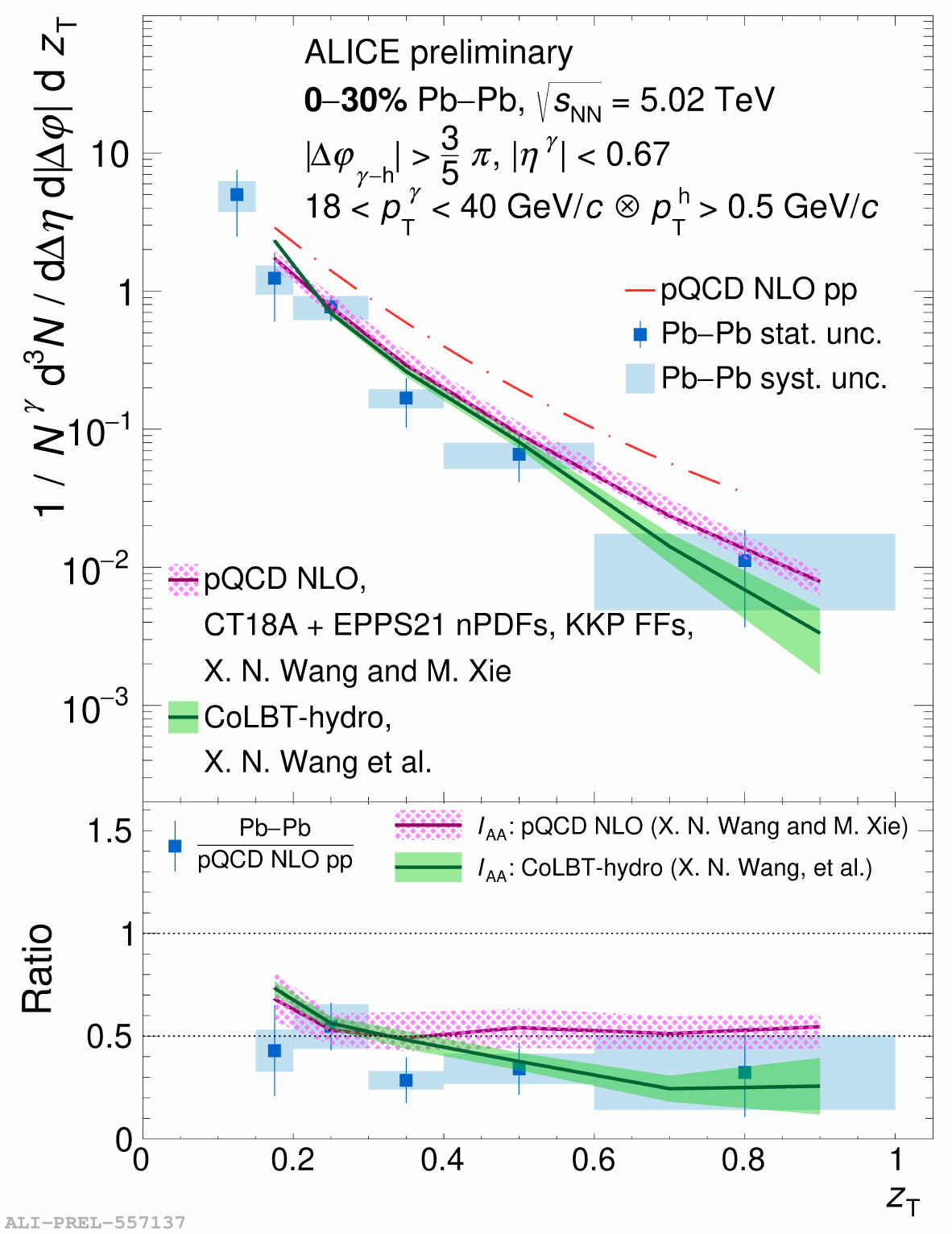}
  \includegraphics[width=4.2cm,clip]{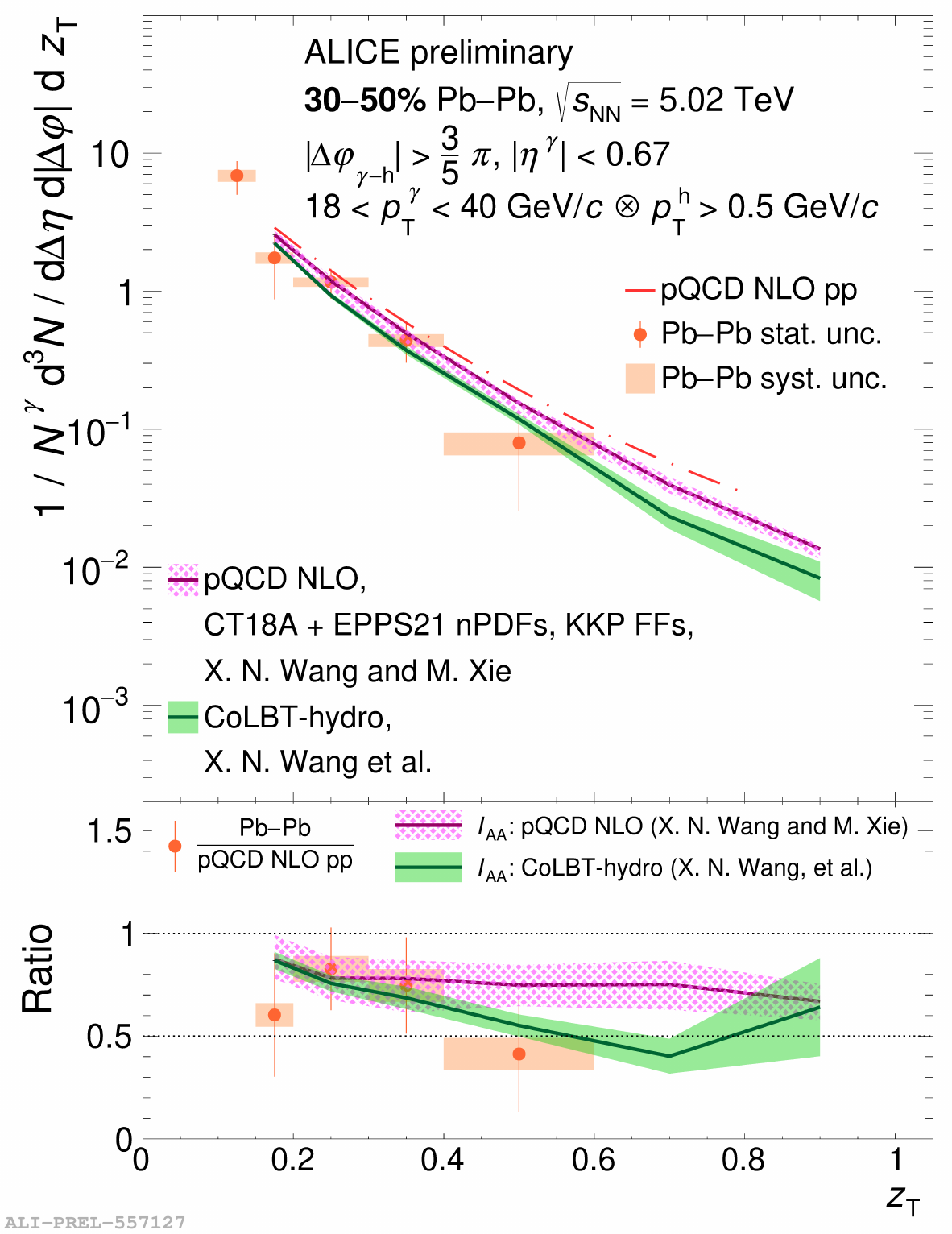}
  \includegraphics[width=4.2cm,clip]{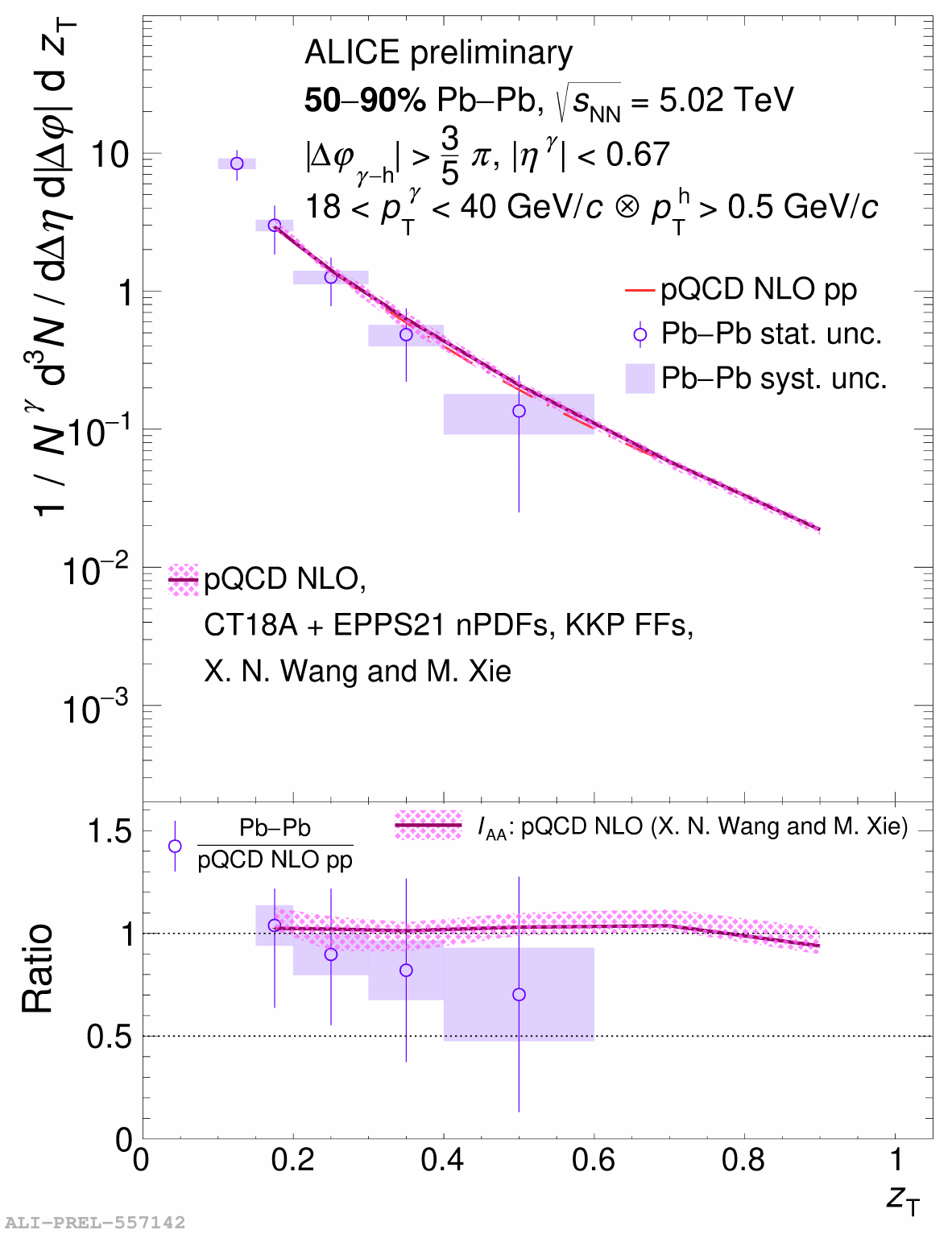}
  \caption{Fragmentation functions $D(z_{\rm T})$ (top panels) and ratios of $D(z_{\rm T})$ in Pb--Pb collisions to pQCD calculations (bottom panels) as a function of \zT.}
  \label{fig-2}    
\end{figure}

 Correlations between \isogamma~and hadrons from the parton fragmentation can give access to the measurement of the jet fragmentation function and the study of any modifications induced by the QGP.
The isolated photons with a $p_{\rm T}^{\gamma}$ between 18 and 40 GeV/$c$ are correlated with hadrons with $p_{\rm T}^{\rm h}$ above 0.5 GeV/$c$. The angular correlation $\Delta\varphi=(\varphi^{\gamma}-\varphi^{\rm h})$ is estimated in \zT=$p_{\rm T}^{\rm h}/p_{\rm T}^{\gamma}$ intervals.
The combinatorial background that affects the angular distribution is subtracted using the mixed event technique. The residual background correlations, triggered by $\pi^{0}$, are removed using the purity ($\it{P}$) correction: $\Delta\varphi\,(\gamma^{\rm iso}) = \frac{1}{P}[ \Delta\varphi\,(N^{\rm iso}_{\rm narrow}) - (1-\it{P})\cdot \rm \Delta\varphi\,(\it{N}^{\rm iso}_{\rm wide})]$. At these energies most of the two photons from $\pi^{0}$ decays merge into a single cluster that populate the $N^{\rm iso}_{\rm wide}$ sample and can be used to estimate the photon decay background in the clusters narrow region.
The integral of the azimuthal correlation distribution with $\Delta\varphi$\,>\,3/5$\,\pi$ is computed for every \zT~bin and the $D(z_{\rm T}) = 1\,/\,N^{\gamma}\,\rm{d}^{3}N\,/\,\rm{d}\Delta\eta\,\rm{d}|\Delta\varphi|\,d \it{z}_{\rm{T}}$ distributions are obtained, as shown in Figure~\ref{fig-2}.
The results are compared with pQCD NLO calculations \cite{PhysRevC.103.034911, PhysRevLett.103.032302} and in central Pb--Pb collisions they are also compared with CoLBT-hydro model \cite{Chen:2017zte}. Good agreement is observed between data and theory, but the discrimination between the two models is not possible yet due to the current uncertainties.
The pQCD NLO calculations for pp collisions have been used as our reference since the measurement in pp was not available: a clear difference can be seen in data with respect to the pQCD NLO calculation and we quantify this variation via the ratio of data to the pQCD reference, shown in the bottom panel of each plot in Figure~\ref{fig-2}. The ratios are below 1 as expected due to the QGP modification and more suppression is visible in central collisions with respect to peripheral. The \zT~distributions are suppressed from central to peripheral collisions and compared to pQCD calculations, reflection of the hadrons suppression at high \pT as shown in the hadrons $R_{\rm AA}$ \cite{ALICE:2012mj}. The ratios are compared with the $I_{\rm AA}$ from theoretical models and are consistent with uncertainties. These measurements have also been compared with related results done at LHC by CMS \cite{CMS:2021otx, CMS:2018mqn} and at RHIC by STAR \cite{STAR:2016jdz} and PHENIX \cite{PHENIX:2012aba}. The trends and magnitude of the ratio are found to be similar. These comparisons are not shown in the proceeding, but they can be found in the presentation.

\section{Conclusions}
We present the measurement of the \isogamma~spectrum and $R_{\text{AA}}$ using  two isolation cone radii $R$ in pp and Pb--Pb collisions by ALICE. These results are compatible with JETPHOX pQCD NLO theory calculations and extend the LHC measurements towards lower \pT. We also report on \isogamma--hadron correlations in Pb--Pb collisions. A centrality-dependent medium modification is seen for the $D(\zT)$ and the results are compatible with models.

%
\bibliography{bibliography}
%
%
%
%

\end{document}